\documentclass[conference]{IEEEtran}
\usepackage{amsmath,amssymb,times}
\usepackage{multirow,epsfig,cite,psfrag}
\usepackage{times, amsmath}
\usepackage{stfloats}
\usepackage{dsfont}
\usepackage{color}
\usepackage{enumerate}
\usepackage{amsfonts}
\usepackage{amsthm}
\usepackage{amsmath}
\usepackage{array}
\usepackage{bbm}
\usepackage{cite}
\usepackage{dsfont}
\usepackage{epsfig}
\usepackage{float}
\usepackage[T1]{fontenc}
\usepackage{graphicx}
\usepackage{indentfirst}
\usepackage{multicol}
\usepackage{subfigure}
\usepackage{url}
\usepackage{multirow}
\usepackage{color}
\usepackage{epstopdf}
\usepackage{algpseudocode}
\usepackage[normalem]{ulem}
\usepackage{float}
\usepackage[nolist]{acronym}
\usepackage{booktabs}
\usepackage{subfigure}
\usepackage{graphicx}

\usepackage{graphicx}
\usepackage{caption}
\usepackage{algorithm}
\usepackage{algpseudocode}
\begin{document}
\title{Opportunistic Scheduling of Machine Type Communications as Underlay to Cellular Networks}
\author{ Samad Ali, Nandana Rajatheva\\%
%\vspace{0.3em} %\\%
Centre for Wireless Communications, University of Oulu, Oulu, Finland\\
samad.ali@oulu.fi, rrajathe@ee.oulu.fi \\
}
\maketitle
\vspace{-2cm}
\begin{abstract}
In this paper we present a simple method to exploit the diversity of interference in heterogenous wireless communication systems with large number of machine-type-devices (MTD). We consider a system with a machine-type-aggregator (MTA) as underlay to cellular network with a multi antenna base station (BS). Cellular users share uplink radio resources with MTDs. Handling the interference from MTDs on the BS is the focus of this article. Our method takes advantage of received interference diversity on BS at each time on each resource block and allocates the radio resources to the MTD with the minimum interference on the BS. In this method, BS does not need to take the interference from MTD into account in the design of the receive beamformer for uplink cellular user, hence, the degrees of freedom is not used for interference management. Our simulation results show that each resource block can be shared between a cellular user and an MTD, with almost no harmful interference on the cellular user.
\end{abstract}

\section{Introduction} \label{sec:introduction}
Conventional wireless communications systems are designed with the goal of providing high data rates for human type users. With the introduction of Internet-of-Things (IoT), wireless networks should provide a new type of connectivity, which is known as machine-to-machine (M2M) communications or machine-type-communications (MTC). MTC is an essential part of the development of the fifth generation (5G) cellular networks \cite{alliance2015ngmn}. Challenges in MTC are different compared to human type traffic. These differences arise from the type of services that require connectivity and the type of communicated data. In human type communication, applications require large data exchanges such as multimedia services or web browsing. Whereas in MTC, applications such as smart metering (electricity, gas, etc.), control and monitoring and public safety should be supported \cite{shariatmadariMagazie}.  From the data type point of view, MTC traffic is comprised of short packet transmissions and the challenge in design of wireless systems for MTC mainly originates from short packet nature of  MTC traffic and different quality-of-service (QoS) requirements of IoT applications. Some IoT services such as public safety and factory automation have stringent reliability and latency requirements \cite{factoryAutomation}, these are known as ultra-reliable MTC (uMTC). Meanwhile, there is a need to design systems that can provide connectivity for a massive number of machine- type-devices (MTD) \cite{MassiveM2M}, known as massive MTC (mMTC). An overview on recent advancements of 3GPP on MTC is given in \cite{3GPPM2M} and current status and future directions are presented in \cite{shariatmadariMagazie}. Examples of narrowband cellular solutions which allocate a part of wireless spectrum only for MTC are Narrowband IoT (NB-IoT) \cite{3GPP-NB-IoT} and Narrowband LTE-M \cite{NBLTE-M}. With a massive number of devices trying to access the network, novel solutions should be developed to increase the spectral efficiency of the wireless communication systems.

In this paper, we focus on providing solutions for massive number of MTDs.  One of the methods of providing connectivity for MTC is by capillary networks \cite{CapillaryMain}, \cite{CapillaryHamid}. In capillary networks, MTC nodes transmit data to a machine-type-aggregator (MTA) and then MTA forwards collected data to the cellular network. In this paper, we assume that MTC node to MTA communication link is an underlay link to cellular communications. This  means that the link will use licensed spectrum, as opposed to other works which assume MTD to MTA linkto be in the unlicensed spectrum. This underlay link shares radio resources with human type cellular users. This causes interference to the main cellular communications and mitigating this interference is the the objective of this paper. Our proposed method is similar to communicating MTC traffic over Femtocells \cite{5GDenseNet}.

We present the idea of opportunistic spatial orthogonozalization (OSO) of MTC  interference on the base station (BS). The concept of OSO takes advantage of the spatial diversity of received interference in the network. This scheme was initially introduced in \cite{shen2009dynamic}, \cite{shen2011opportunistic} and \cite{IntDraining} for cognitive radio networks, where there are a large number of secondary user transmit candidates. In each time slot, one of the secondary users is selected for transmission, such that interference from selected secondary transmitter is minimum after being multiplied by receive beamformer of the primary link. The ideal scenario in OSO is that interference falls into the null space of the primary receiver's beamformer, and hence, there is no interference on the primary link. The main requirements for this technique are knowledge of channel-state-information (CSI) and the existence of a large number of secondary transmit candidates. In our scenario, we assume that MTDs do not transmit delay sensitive data and it is possible for them to wait, in case that there is no opportunity to transmit. This method is further developed in other publications for multiple-input multiple-outpu (MIMO) interference channels in \cite{perlaza2010spectrum}, \cite{junghoon2010}, \cite{lee2011interference}. In MIMO this concept is modified and the secondary user transmits data in the direction of unused singular values of primary transmit precoder and decoder. A similar approach is taken in \cite{yang2013opportunistic} for intercell-interference mitigation. The idea of transmitting in the null-space of primary link is proposed for interference mitigation between macro and small cell in heterogenous networks \cite{Oulu2014draining}. The concept of OSO in conventional systems has some drawbacks that can be listed as follows:
\begin{itemize}
	\item Considering a very large number of secondary users is not realistic in normal human type communications.
	\item Receive beamformer of the primary link changes at each time step. This means that the secondary user's opportunity to transmit can be lost in the next time step and it might not be able to transmit its data.
\end{itemize}

In our scenario, the above mentioned drawbacks do not exist. First, the assumption of very large number of users is realistic since our solution is for mMTC. Second, MTDs transmit small packets and even a small transmit opportunity is enough for the MTD to transmit its data.

The rest of the paper is organized as follows. In section \ref{sysmodel} we present the system model, signal and SINR equations. In section \ref{problem} the proposed method is developed and presented in details. Simulation results are presented in section \ref{results} and conclusions are drawn in section \ref{conclusions}.

\section{System Model}\label{sysmodel}
\begin{figure}[!h]
 \begin{center}
   \includegraphics[width=0.45\textwidth]{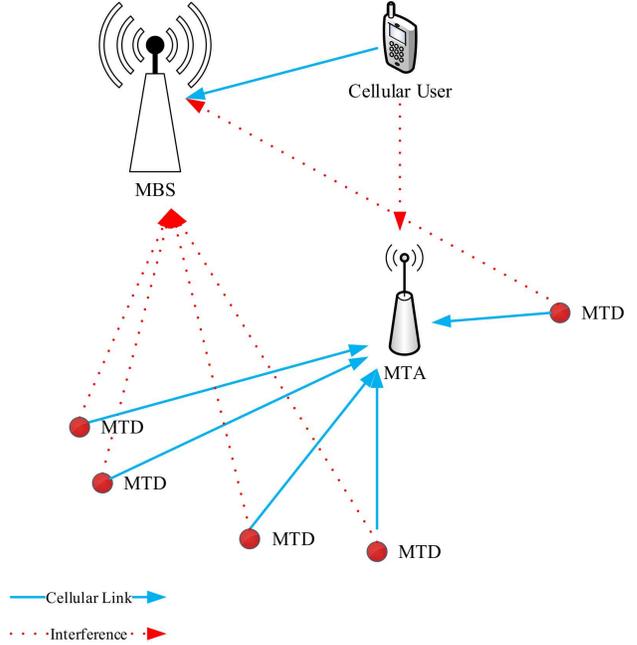}
 \end{center}
 \caption{System Model}
 \label{fig:sysmodel2}
\end{figure}

We consider the uplink of a single cell system with a BS and an underlay machine-type-aggregator (MTA). Cellular user, MTDs and MTA are considered to be single antenna devices. BS has $\rm{M}$ antennas and the receive beamformer in BS is $\mathbf{w_{\rm c}} \in \mathbb{C}^M $, which can be selected by different receive methods, for example, in case of one uplink user, it is selected to be maximal ratio combining (MRC). The channel between BS and cellular users is $\mathbf{h_{\rm c}} \in \mathbb{C}^M $ which is acquired at BS at each transmission period. There are $\rm{K}$ MTDs that communicate with the MTA, using the same radio resources as uplink of BS. Furthermore, we assume that MTDs are stationary or low mobility users, with channel from MTD $\rm{k}$ to the the BS shown by $\mathbf{h_{\rm kb}} \in \mathbb{C}^M $. At each transmission time BS carries out the beamformer design $\mathbf{w_{\rm c}}$. BS performs radio resource allocation for the uplink users to communicate with BS and MTDs to communicate with MTA. Considering the beamformer $\mathbf{w_{\rm c}}$, for each resource block, BS selects one MTD, $\rm{k} \in \mathcal{K} = \{1,...,K\}$, such that received singal to interference plus noise (SINR) of uplink user is maximized. The system model is given in Fig \ref{fig:sysmodel2}.

The received signal at the BS $\mathbf y_{\rm bs}$ from cellular user is given by:

\begin{equation}
\begin{aligned}\label{eq:RUplink}
&{\mathbf y}_{\rm bs}& = {\mathbf h}_{\rm c}x_{\rm c} + {\mathbf h}_{\rm kb}x_{\rm m} + {\mathbf n}_{\rm bs}\\
\end{aligned}
\end{equation}

where  $x_{\rm c}$ and $x_{\rm m}$ are the transmit symbols of the cellular user and MTD respectively. $\mathbf{h_{\rm c}} \in \mathbb{C}^M $ is channel between cellular user and BS. MTD is selected from $\rm{K}$ devices with channels $\mathbf{h}_{\rm kb} \in \{\mathbf{h}_{\rm 1b},...,\mathbf{h}_{\rm Kb} \} $. White Gaussian noise is represented by ${\mathbf n}_{\rm bs}$.

After applying the receive beamformer at BS, receiver has the following signal:
\begin{equation}
\begin{aligned}\label{eq:RUplinkbf}
&{\hat{\mathbf y}}_{\rm bs}& = \mathbf{w_{\rm c}}{\mathbf h}_{\rm c}x_{\rm c} + \mathbf{w_{\rm c}}{\mathbf h}_{\rm kb}x_{\rm m} + \mathbf{w_{\rm c}}N_0\\
\end{aligned}
\end{equation}

From this equation, now we can write the SINR of the cellular user at BS:
\begin{equation}
\begin{aligned}\label{eq:MBSSINR}
\gamma_{\rm c} =  \frac{P_c |\mathbf{w_{\rm c}}{\mathbf h}_{\rm c}|^2}{P_k| \mathbf{w_{\rm c}}{\mathbf h}_{\rm kb} |^2 + | \mathbf{w_{\rm c}} |^2N_0 }
\end{aligned}
\end{equation}

In a similar manner the received signal at the MTA can be written as follows:
\begin{equation}
\begin{aligned}\label{eq:RMTG}
y_{\rm m} = {h}_{\rm k}x_{\rm k} + {h}_{\rm cs}x_{\rm c} + {n}_{\rm s}\\
\end{aligned}
\end{equation}

Here ${h}_{\rm k}$ is the channel between MTD and MTA, and ${h}_{\rm cs}$ is the channel between cellular user and MTA. Additive white Gaussian noise (AWGN) is MTA is ${n}_{\rm s}$.  The received SINR at MTA can be given by:
\begin{equation}
\begin{aligned}\label{eq:scSINR}
\gamma_{\rm k} =  \frac{P_k|{h}_{\rm k}  |^2}{P_c| {h}_{\rm cs} |^2 + N_0 }
\end{aligned}
\end{equation}
$P_k$ and $P_c$ are transmit powers of MTD and cellular user, respectively. We consider that the cellular user is selected such that the interference can be treated as background noise at the MTA, so the SINR in the equation (\ref{eq:scSINR}) becomes:
\begin{equation}
\begin{aligned}\label{eq:scSINR2}
\gamma_{\rm k} =  \frac{P_k\lVert {h}_{\rm k}  \lVert^2}{I_0 + N_0 }
\end{aligned}
\end{equation}

The cellular user selection can be performed by a location based interference limiting method such as \cite{ILA}. Considering the maximum transmit power allowed for cellular user, the users that are far enough from MTA are opted for resource sharing. By using this method, we only focus on interference minimization on BS and treat the interference at MTA as background noise.

\section{Proposed Scheme}\label{problem}
We assume an OFDMA based system such as LTE where radio resources are divided into orthogonal sub-units in time and frequency  domain. In LTE, each resource block has $180 $ kHz bandwidth and one millisecond duration. Several resource blocks in the frequency domain can be allocated to one cellular user at a given time. We assume that one resource block is shared between uplink cellular user and one MTD to MTA link. As mentioned earlier, the objective of this paper is to minimize the interference on the BS. We assume that the BS has \textit{a priori} information about the devices that have data to transmit. This assumption is valid in scenarios such as periodic meter reporting. Since the locations of the MTD nodes are fixed, we also assume that the BS has channel state information of the MTD to BS links \cite{dhillon2014fundamentals}. Once the BS applies our proposed scheme and allocates radio resources for an MTD, it can send the transmission permission using a signaling scheme called fast uplink grant \cite{LTE14outlook}. In fast uplink grant, users skip the scheduling requests and receive uplink grant periodically. Combination of fast uplink grant and our proposed OSO provide two benefits in MTC systems. One is reducing the signalling overhead and random access channel congestion, and the second is increased spectral efficiency by providing connectivity to the MTD user with almost no extra cost in form of radio resources.

\subsection{Minimum Interference on BS}\label{scheme}
In this section, we present the solution for selecting one user for each time frequency resource block. If we assume that BS has CSI knowledge of the MTD - BS links, at each transmission period, the user which causes minimum interference at the BS is selected for resource sharing with cellular user.  This is equal to the minimization of the term $P_k |\mathbf{w_{\rm c}}{\mathbf h}_{\rm kb} |^2$  which is a fairly easy task to do. $P_c$ can be a constant value, or defined by a transmit power control algorithm. If we assume that the channel between MT and MTA is not known, MT devices can transmit with a predefined maximum transmit power. BS in this stage simply selects one of the MTDs that causes the minimum interference on the BS. With this method applied, the SINR equation for cellular user in BS is:

\begin{equation}
\begin{aligned}\label{eq:MBSSINROSO}
\gamma_{\rm c} =  \frac{P_c |\mathbf{w_{\rm c}}{\mathbf h}_{\rm c}|^2}{min_{i=1,...K}P_i| \mathbf{w_{\rm c}}{\mathbf h}_{\rm ib} |^2 + | \mathbf{w_{\rm c}} |^2N_0 }
\end{aligned}
\end{equation}

If we assume MRC receiver at BS \cite{tse2005fundamentals}, $\mathbf{w_{\rm c}} = \mathbf{h_{\rm c}}^H$ and assuming MTD $\rm k$ as the one with minimum interference, we can have the following equation for SINR:
\begin{equation}
\begin{aligned}\label{eq:MBSSINROSOMRC}
\gamma_{\rm c} =  \frac{P_c \lVert{\mathbf h}_{\rm c}\lVert^2}{\frac{P_k| \mathbf{h_{\rm c}}^H{\mathbf h}_{\rm kb} |^2}{\lVert{\mathbf h}_{\rm c}\lVert^2}  + N_0 }
\end{aligned}
\end{equation}

To further evaluate the performance of this scenario, we can write the probability of outage for the cellular user:

\begin{equation}
\begin{aligned}\label{eq:pout}
P_{\rm out} =  Pr\{ \frac{P_c \lVert{\mathbf h}_{\rm c}\lVert^2}{\frac{P_k| \mathbf{h_{\rm c}}^H{\mathbf h}_{\rm kb} |^2}{\lVert{\mathbf h}_{\rm c}\lVert^2}  + N_0} \leq \delta_{th} \}
\end{aligned}
\end{equation}
where $\delta_{th}$ is maximum allowed outage probability. To avoid outage, the interference term should be smaller than a threshold, show by $\delta_{int}$:

\begin{equation}\label{eq:QoS}
\begin{aligned}
\frac{P_k| \mathbf{h_{\rm c}}^H{\mathbf h}_{\rm kb} |^2}{\lVert{\mathbf h}_{\rm c}\lVert^2} < \delta_{i}
\end{aligned}
\end{equation}
If the above criteria is satisfied, the selected MTD $k$ can transmit data to the MTA without harmful interference.

The probability to find an MTD that will satisfy the criteria in Eq. \ref{eq:QoS} becomes one when $K \rightarrow \infty$. To prove this argument, we assume that $\frac{| \mathbf{h_{\rm c}}^H{\mathbf h}_{\rm ib} |^2}{\lVert{\mathbf h}_{\rm c}\lVert^2}, i=1...K$ are random variables $X_1,...,X_K$ and $X_{min}$ is the random variable defined as $\min_i\{X_1,...,X_K$\}. The probability that $X_{min}$ is smaller than a defined threshold  $\delta_{i}$ is:
\begin{align}\label{probobo}
P(X_{min} < \delta_{i}) & = 1-P(X_{min} > \delta_{i}) \nonumber \\
& = 1-\int P(X_{1} > \delta_{i},..,X_{K}  > \delta_{i}| \mathbf{h}_c) f(\mathbf{h_c})d\mathbf{h}_c \nonumber\\
 & = 1 - \int (1 - \Phi(\delta_{i}))^K f(\mathbf{h_c})d\mathbf{h}_c \nonumber\\
\end{align}
 where $\Phi(\delta_{i})$ is the \textsc{cdf} of the probability that all of the random variables are smaller than $\delta_{i}$. Since $\delta_{i} \in \mathbb R $ and $\Phi(\delta_{min}) \in (0,1)$ the term $(1 - \Phi(\delta_{min})^K) = 0$ as  $K \rightarrow \infty$, hence, $P(X_{min} < \delta_{min}) = 1$.

\subsection{Allocating more than one radio resource block}\label{mutipleRBs}
We discussed the method for utilizing one radio resource block, selecting one MTD node to transmit. In can further be developed for a group of radio resource blocks. If more than one LTE resource block is allocated for cellular user with a frequency selective channel (we assume frequency flat fading in each resource block), then the beamformer $\mathbf{w}_{\rm c}$ is different for each resource block, and more than one MT device can be selected for radio resource sharing with the cellular user. In this case, the problem becomes more complicated since we do not want to allocate more than one radio resource block for the same MTD. If we assume that there are $N$ radio resource blocks available at each time for sharing, then we need to select $N$ out of $K$ MTDs. For each subcarrier, MTD that causes the minimum interference is selected.

We from the matrix from channels of MTDs to BS with $\mathbf{H}_{\rm int} = [\mathbf{h}_{\rm 1b},...,\mathbf{h}_{\rm Kb}] \in \mathbb{C}^{N\times M}$  and matrix of beamformer as $\mathbf{W}_{\rm c} = [\mathbf{w}_{\rm c1},...,\mathbf{w}_{\rm cN}]^T \in \mathbb{C}^{M\times K}$. We denote the matrix of interferences as $\mathbf{I}_{\rm int} = [\mathbf{i}_{\rm 1},...,\mathbf{i}_{\rm N}]^T\in \mathbb{C}^{N\times K}$ in which $\mathbf{i}_{\rm n} \in \mathbb{C}^{K}$ is the vector of interferences on the BS from all MTDs on resource block $\rm n$. It is clear that $\mathbf{I}_{\rm int}  =  \mathbf{W}\mathbf{H}$. The objective here is to find the user with the minimum interference on each resource block, and also assign this resource block to only one MTD. For this, we propose the following simple algorithm.

\begin{algorithm}[H]
  \caption{Matching interferences and beamformers}
       \begin{algorithmic}[1]
        \State Calculate matrix $\mathbf{I}_{\rm int}$ and for each row $\mathbf{i}$ find the column index of smallest element
        \State\label{2}Compare the index of minimum MT devices for all of them, if each index is unique, go to  \ref{end}
        \State If one index is repeated more than one time, then match the beamformer that receives smaller interference and for the other one, select the second smallest MT devices
        \State Repeat \ref{2} for second smallest MT device, until End.
        \State\label{end}End
       \end{algorithmic}
       \label{MatchingAlgorithm}
        \end{algorithm}

With this algorithm, we enable radio resource sharing between cellular users and MTDs almost for all of the radio resource blocks in the system.

\section{Simulation Results}\label{results}
In this section numerical results are presented to validate the effectiveness of the proposed method. We use LTE parameters in which each resource block is $180$ Khz in frequency domain and one millisecond in time domain. Resource blocks are primarily allocated to cellular users and CSI is acquired at the BS to design the beamformer for the cellular user. For each cellular user, in frequency domain more than one resource block can be allocated depending on the required data rate. Assuming that the coherence bandwidth of the uplink channel for cellular user is approximately $180$ kHz, each resource block has different fading parameters and the receive beamformer $\mathbf{w}_{\rm c}$ is different for each resource block. We fix the location of MTD users, but the cellular user has a different location at each time step. Simulation parameters are given in Table \ref{tab:simparameters}.

\begin{table}[!h]
\caption{Simulation Parameters} %title of the table
\centering % centering table
\begin{tabular}{c|c} % creating eight columns
\hline\hline %inserting double-line
Parameter & Value\\
\hline\hline % inserts single-line
Number of Antennas in BS (M) & 4\\
\hline
Cell Radius & 500 m\\ % Entering row contents
\hline
MTA Radius & 250 m\\
\hline
Number of resource blocks & 20\\
\hline
Noise Figure at BS and MTA & 2 dB\\
\hline
CU to BS Path Loss Model  & 128.1 + 36.7log(d[km])\\
\hline
CU Target SINR & 10 dB\\
\hline
Noise spectral density & -174 dBm/Hz\\
\hline
\end{tabular}
\label{tab:simparameters}
\end{table}

\begin{figure}[!h]
 \begin{center}
   \includegraphics[width=0.52\textwidth]{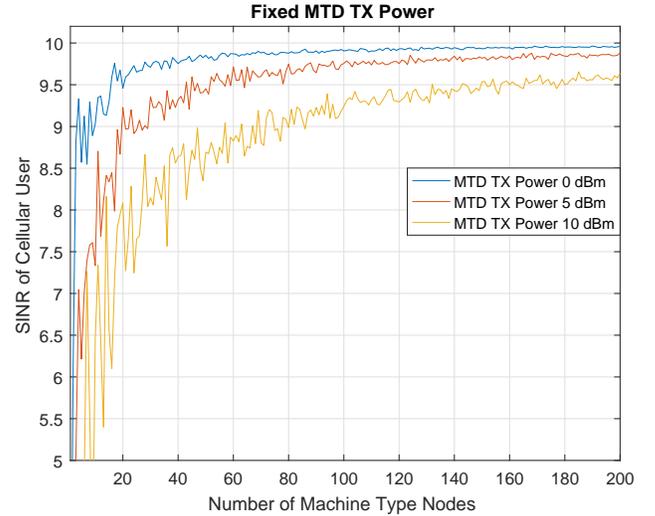}
 \end{center}
 \caption{SINR of the cellular user with fixed MTD TX power}
 \label{fig:result01}
\end{figure}

\begin{figure}[!h]
 \begin{center}
   \includegraphics[width=0.52\textwidth]{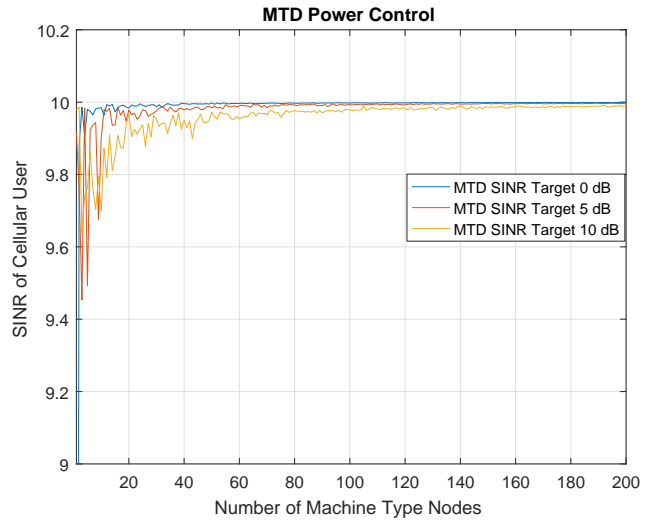}
 \end{center}
 \caption{SINR of the cellular user with MTD TX power control}
 \label{fig:result02}
\end{figure}

First, we present the results for a scenario with only one resource block to study the possible effect of number of MTDs on the SINR of cellular user. For one resource block two transmit power strategies for MTDs are considered. First one is a fixed transmit power from MTDs, and second is a transmit power control to satisfy the required SNR at MTA. Since BS has the knowledge of the location of MTDs, it is rational to assume that power control by MTDs can be approximated by BS. Assuming that target SINR for cellular user is $10$ dB and sharing the radio resource between MTDs and cellular users, the degradation of SINR of cellular userdue to MTD interference, is presented in Fig \ref{fig:result01} for the case with fixed transmit power for MTDs, and in Fig \ref{fig:result02} for the scenario of power control by MTDs.
In Fig \ref{fig:result01} the results show the effect of interference from MTD on SINR of cellular user for a different number of MTDs. This figure gives a clear example of how a  large number of MTDs provides the diversity to select one MTD with very small interference on BS. For transmit power of $0$ dBm, this interference almost has not effect on cellular user SINR.

In Fig \ref{fig:result02} a similar plot is given for the scenario of having power control for MTDs. Since distances among MTDs and MTA are small, MTDs transmit with lower powers which leads to less interference. However,l if the number of MTDs is small, it can cause high interference on cellular user. With a large number of MTDs, there is no interference on cellular user.

\begin{figure}[!h]
 \begin{center}
   \includegraphics[width=0.52\textwidth]{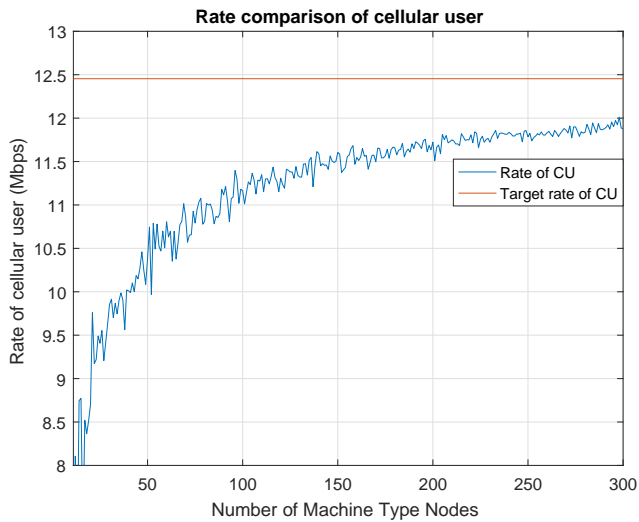}
 \end{center}
 \caption{Throughput of the system}
 \label{fig:result03}
\end{figure}

For the scenario presented in section \ref{mutipleRBs} we present the throughput of cellular user with resource sharing and the target rate with SINR of $10$ dB in Fig \ref{fig:result03}. We assume that $20$ resource blocks are available in the frequency domain and they are shared with MTDs. If no resource sharing is done, the target rate line is the throughput of the cellular user. When the resources are shared with 20 MTDs, this throughput is lower if there is no method of interference management. After applying our scheme and increasing the number of MTDs, one can see that the throughput of the cellular user gets closer to the rate without interference. This means that a total of 20 MTDs are also transmitting at almost no extra cost in in spectral domain. Since each of these MTDs are transmitting in a different frequency channel, their data can be independently decoded in MTA.

\section{Conclusions}\label{conclusions}
In this paper, we presented a method for coexistence of MTC and human type communications in cellular networks by exploiting the diversity of MTD interference. This scheme is similar to multi-user diversity, with the difference that the spatial diversity of interference is exploited. This diversity is available due to the very large number of MTDs in the system. Our proposed scheme enables higher spectral efficiency by sharing radio resources and higher power efficiency by short range transmissions of the MTDs to MTA. Our results validate that a large number of MTDs can be taken advantage of to allocate radio resources for MTC traffic almost at no extra cost in terms of radio resources or spatial degrees of freedom.

\section{Acknowledgement}
This research is supported by P2P-SMARTEST project, Peer to Peer Smart Energy Distribution Networks (http://www.p2psmartest-h2020.eu/), an Innovation Action funded by the H2020 Programme, contract number 646469.
\bibliographystyle{IEEEtran}

\bibliography{diIntDiv}

\end{document}